\documentclass[
  ,final            
  ]
  {aipproc}

\layoutstyle{8x11double}
\usepackage{graphicx,color}
\usepackage{widetext}

\begin{document}

\title{Light scalar susceptibilities and the $\pi^0-\eta$ mixing}

\author{Ricardo Torres Andr\'es}{
  address={Departamento de
F\'{\i}sica Te\'orica II. Univ. Complutense. 28040 Madrid. Spain.}
}

\author{\'Angel G\'omez Nicola}{
  address={Departamento de
F\'{\i}sica Te\'orica II. Univ. Complutense. 28040 Madrid. Spain.}
}

\begin{abstract}
 We have performed a thermal analysis of the light scalar susceptibilities in the context of SU(3)-Chiral Perturbation Theory to one loop taking into account the QCD source of isospin breaking (IB), i.e corrections coming from $m_u\neq m_d$. We find that the value of the connected scalar susceptibility in the infrared regime and below the critical temperature is entirely dominated by the $\pi^0-\eta$ mixing, which leads to model-independent $\mathcal{O}(\epsilon^0)$ corrections, where $\epsilon\sim m_d-m_u$, in the combination $\chi_{uu}-\chi_{ud}$ of flavour breaking susceptibilities.
\end{abstract}

\maketitle

\section{Introduction}

The low-energy sector of QCD has been successfully described within the chiral lagrangian framework. Chiral Perturbation Theory (ChPT) is based on the spontaneous breaking of chiral symmetry and provides a consistent, systematic and model-independent scheme to calculate low-energy observables \cite{we79,Gasser:1983yg,Gasser:1984gg}. This formalism has been also extended to include finite temperature effects, in order to describe meson gases and their evolution towards chiral symmetry restoration \cite{Gasser:1986vb,Gerber:1988tt}. The effective ChPT lagrangian is constructed as an expansion of the form ${\cal L}={\cal L}_{p^2}+{\cal L}_{p^4}+\dots$ where $p$ denotes a meson momentum or mass compared to the chiral scale $\Lambda_{\chi}\sim 4\pi F\!\!\simeq 1$ GeV where $F$ is the pion decay constant in the chiral limit. Each term of the expansion is accompanied by a low energy constant (LEC) which has to be determined experimentally.

ChPT can take into account both QCD (due to the light quark masses diference $m_u-m_d\neq0$) and electromagnetic IB by means of new terms that implement the chiral symmetry breaking pattern. The former generates a $\pi^0-\eta$ mixing in the $SU(3)$ lagrangian which introduces corrections of order $(m_d-m_u)/m_s$ which will be important when considering some combinations of the light scalar susceptibilities at finite temperature. On the other hand, the presence of electromagnetic interactions induces mass differences for the light mesons through the presence of virtual photons. These corrections have been included in the ChPT effective lagrangian \cite{Ecker:1988te,Urech:1994hd} by means of terms like ${\cal L}_{e^2}$, ${\cal L}_{e^2p^2}$ and so on, with $e$ the electric charge. These terms are easily incorporated in the ChPT power counting scheme by considering formally $e^2=\mathcal{O}(p^2\slash F^2)$.

The aim of this work is to explore within the thermal ChPT formalism the IB corrections to the next-to-leading quark condensates and their corresponding light scalar susceptibilities, both physical objects being directly related to chiral symmetry restoration. More details can be found in \cite{agnrtathermal}.

\section{Light quark condensates to one loop}

We  have calculated to one loop the light quark condensates $\langle\bar uu\rangle$ and $\langle\bar dd\rangle$ in SU(3)-ChPT taking into account both sources of IB. The main distinctive feature with respect to $SU(2)$-ChPT calculations  is that, in this case, as commented above, a $\pi^0-\eta$ mixing term appears through the tree-level mixing angle $\varepsilon$ defined by $\tan 2\varepsilon=\frac{\sqrt{3}}{2}\frac{m_d-m_u}{m_s-\hat m}$. The sum and difference of quark condensates are

\begin{widetext}
\begin{equation}
\label{condsu3sum}
  \langle \bar u u + \bar d d \rangle_T^{(3)}=\langle \bar u u + \bar d d \rangle_0^{(3)}+2F^2B_0\bigg(\frac{1}{3}\left(3-\sin^2 \varepsilon\right) g_{\pi^0}(T)+2g_{\pi^\pm}(T)+g_{K^0}(T)+g_{K^\pm}(T)+\frac{1}{3}\left(1+\sin^2 \varepsilon\right)g_\eta(T)\bigg)
\end{equation}

\begin{equation}
\label{condsu3dif}
\langle \bar u u - \bar d d \rangle_T^{(3)}=\langle \bar u u - \bar d d \rangle_0^{(3)}+2F^2B_0\left(\!
\frac{\sin 2\varepsilon}{\sqrt{3}}\left[g_{\pi^0}(T)-g_{\eta}(T)\right]+g_{K^\pm}(T)-g_{K^0}(T)\!\right)
\end{equation}
\end{widetext}

where $B_0=\frac{M^2_\pi}{m_u+m_d}+\mathcal{O}(\epsilon)$, and $$\!g_i(T)\!=\frac{1}{4\pi^2F^2}\int_0^\infty dp \frac{p^2}{E_p} \frac{1}{e^{\beta E_p}-1},$$  with $E_p^2=p^2+M_i^2$ and $\beta=T^{-1}$.
 
The subscript $0$ refers to the zero temperature results, which can be found in \cite{Nicola:2010xt}. As a nontrivial check of our calculation, one can see that the condensates (\ref{condsu3sum})-(\ref{condsu3dif}) are finite and $\mu$-scale independent with the renormalization of the LEC, including
the EM ones, given in \cite{Gasser:1984gg,Urech:1994hd}.

\section{Light scalar susceptibilities and the role of the $\pi^0-\eta$ mixing}

Different light quark masses allow to consider three independent light scalar susceptibilities defined as 
\begin{equation}
\chi_{ij}=-\frac{\partial}{\partial m_i}\langle\bar q_j q_j\rangle=\frac{\partial^2}{\partial m_i\partial m_j} \log Z(m_u\neq m_d).
\label{indepsus}   
\end{equation}

For the sake of simplicity we are setting $e=0$ from now on, since electromagnetic corrections are small and they are not relevant for our present discussion. Then, to leading order in the mixing angle, the  contribution of the $\pi^0-\eta$ mixing in the quark condensate sum (\ref{condsu3sum}) is of order $\epsilon^2$ whereas for (\ref{condsu3dif}) it goes like $\varepsilon$. The thermal functions $g_i(T,M_i),\, i=\pi^0, \eta$ are suppressed by those coefficients and the quark condensates do not receive important corrections. The important point is that differentiating with respect to a light quark mass is essentially the same as differentiating with respect to $\varepsilon\sim \frac{m_d-m_u}{m_s}$,
so the suppression of the thermal functions is smaller in the case of the susceptibilities than in the quark condensate. 

Because of the linearity in $\varepsilon$ of (\ref{condsu3dif}) for a small mixing angle, the combinations $\chi_{uu}-\chi_{ud}$ and $\chi_{dd}-\chi_{du}$ receive an $\mathcal{O}(1)$ IB correction due to $\pi^0-\eta$ mixing, which would not be found if $m_u=m_d$ is taken from the beginning. The analysis of the $\varepsilon$-dependence  of (\ref{condsu3dif}) shows that, up to $\mathcal{O}(\epsilon)$, $\chi_{uu}\simeq \chi_{dd}$, so combinations like $\chi_{uu}-\chi_{dd}$, which also vanish with $m_u=m_d$, are less sensitive to IB.

One can also relate these flavour breaking susceptibilities with the connected and disconnected ones \cite{Smilga:1995qf}, often used in lattice analysis \cite{Ejiri:2009ac,DeTar:2008qi}: $\chi_{dis}=\chi_{ud}$, and $\chi_{con}=\frac{1}{2}(\chi_{uu}+\chi_{dd}-2\chi_{ud})$.
From the previous analysis, we get $\chi_{con}\simeq\chi_{uu}-\chi_{ud}$. 

Therefore, our model-independent analysis including IB effects provides the leading nonzero contribution for the connected susceptibility which arises partially from $\pi^0-\eta$ mixing. This is particularly interesting for the lattice, where artifacts such as taste-breaking mask the behaviour of $\chi_{con}$ with the quark mass and $T$ when approaching the continuum limit \cite{DeTar:2008qi}. 
In fact, our ChPT approach is useful to explore the chiral limit ($m_{u,d}\rightarrow 0$) or infrared (IR) regime, which gives a qualitative picture of the behaviour near chiral symmetry restoration. In this regime $M_\pi\ll T\ll M_K$, and therefore we can neglect thermal heavy particles, which are exponentially suppresed.

The leading order results for the connected and disconnected susceptibilities at zero temperature are the following

\begin{widetext}
\begin{equation}
\label{con}
\chi^{IR}_{con}(T=0)=8B_0^2\left[2L_8^r(\mu)+H_2^r(\mu)\right]-\frac{B_0^2}{16\pi^2}\left(1+\log\frac{M^2_K}{\mu^2}\right)-\frac{B_0^2}{24\pi^2}\log\frac{M^2_\eta}{\mu^2}+\mathcal{O}(\epsilon),
\end{equation}

\begin{equation}
\label{dis}
\chi^{IR}_{dis}(T=0)=32B_0^2L_6^r(\mu)-\frac{3B_0^2}{32\pi^2}\left(1+\log{\frac{M^2_{\pi}}{\mu^2}}\right)+\frac{B^2_0}{288\pi^2}\left(5\log\frac{M^2_\eta}{\mu^2}-1\right)+\mathcal{O}(\epsilon).
\end{equation}
\end{widetext}

The log term of equation (\ref{dis}) is the dominant at $T=0$ and can be found in \cite{Smilga:1995qf}, but the connected IR susceptibility (\ref{con}) is not zero at $T=0$, because it receives contributions of order $\mathcal{O}(1)$ in the mixing angle.

If we consider the pion gas in a thermal bath, then expressions (\ref{con})-(\ref{dis}) are modified according to

\begin{widetext}
\begin{equation}
\label{conT}
\left[\chi_{con}(T)-\chi_{con}(0)\right]^{IR}=\frac{B_0^2}{18}\frac{T^2}{M_\eta^2}+\mathcal{O}\left(\epsilon\, B^2_0\,\frac{T^2}{M^2_\eta}\right)+\mathcal{O}\left(\exp\left[-\frac{M_{\eta, K}}{T}\right]\right),
\end{equation}

\begin{equation}
\label{disT}
\left[\chi_{dis}(T)-\chi_{dis}(0)\right]^{IR}=\frac{3B^2_0}{16\pi}\frac{T}{M_\pi}+\mathcal{O}\left(\epsilon\, B^2_0\,\frac{T^2}{M^2_\eta}\right)+\mathcal{O}\left(\exp\left[-\frac{M_{\eta, K}}{T}\right]\right).
\end{equation}
\end{widetext}

Note that, as we have already mentioned, the eta mass term in equation (\ref{conT}) and in the subleading corrections in the mixing angle comes from the $\epsilon$-analysis and the IR expansion of the $g_1(M_\pi)$, and does not have anything to do with thermal etas.

\begin{figure}
\includegraphics[scale=0.7]{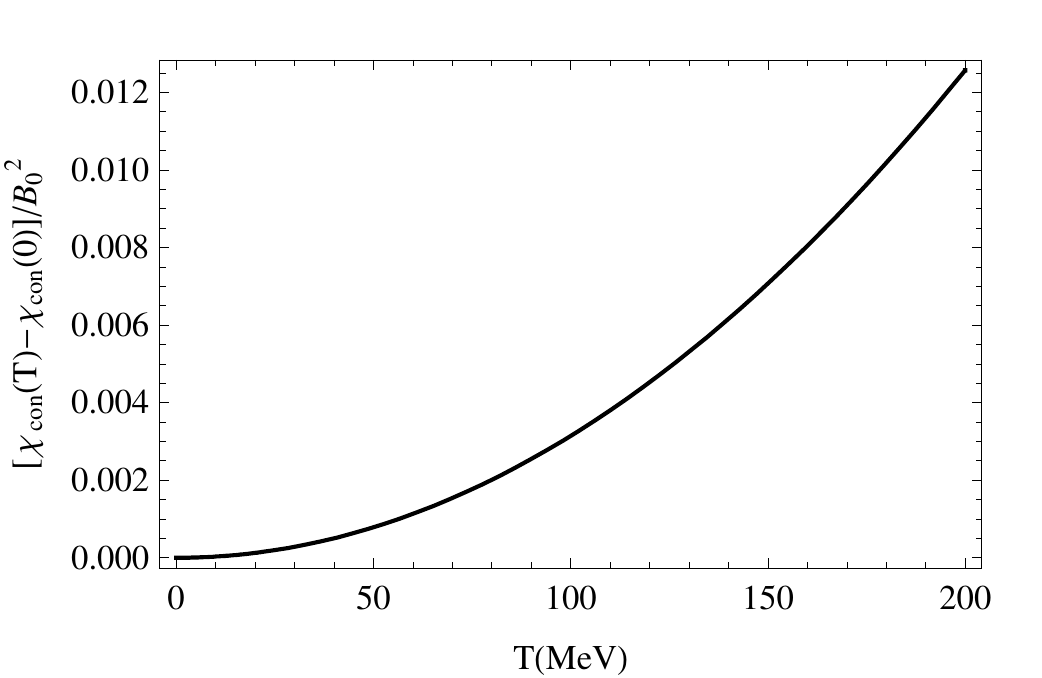}
\label{fig:figa}
\caption{Connected IR susceptibility normalized to $B^2_0$, for fixed tree level eta mass.}
\end{figure}

\begin{figure}
\includegraphics[scale=0.7]{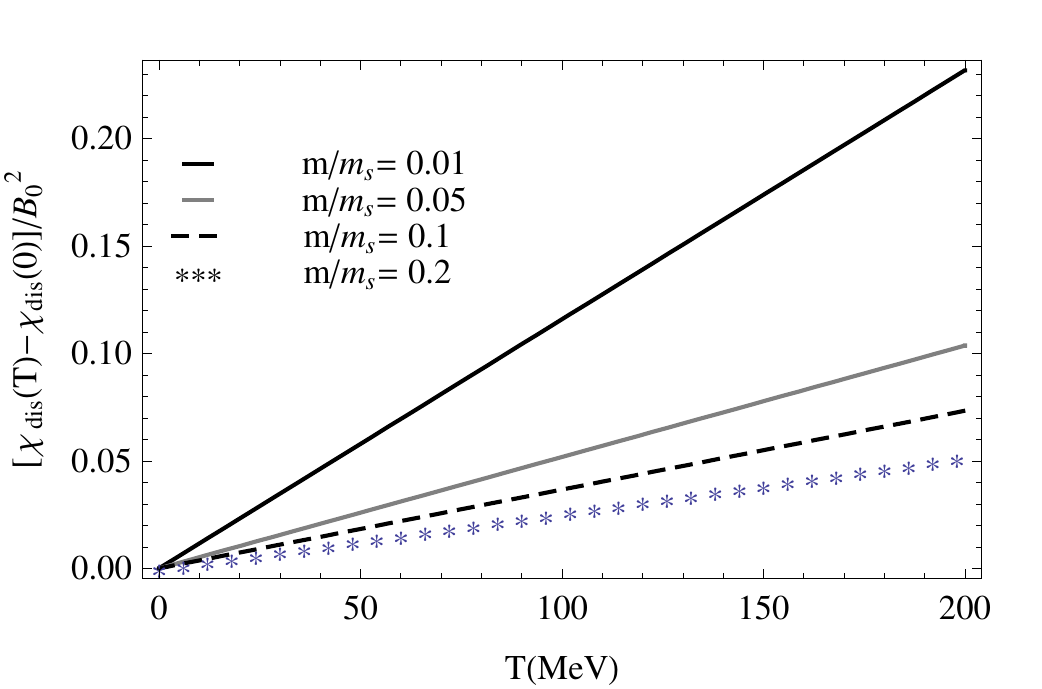}
\label{fig:figb}
\caption{Disconnected IR susceptibility normalized to $B^2_0$, for several light quark mass ratios and fixed tree level eta mass.}
\end{figure}

Figure (\ref{fig:figa}) and (\ref{fig:figb}) show, respectively, the connected susceptibility (\ref{conT}) for fixed tree level eta mass (proportional to $\sqrt{B_0\, m_s}$ in the IR regime), and the disconnected one (\ref{disT}) for several values of the light quark mass ratio $m/m_s$, and also with fixed tree level eta mass.
The leading scaling with $T$ and the light quark mass in this regime for the disconnected piece goes like $\frac{T}{\sqrt{m}}$, i.e the same scaling calculated in \cite{Smilga:1995qf,Ejiri:2009ac}; whereas the connected susceptibility grows quadratic in $T$ over a mass scale much greater than the SU(2) Goldstone boson's one. Therefore, in the continuum limit, we only expect $\chi_{dis}$ to peak near the transition.

\vspace*{4mm}
\begin{theacknowledgments}
 
 R.T.A would like to thank C\' andida Garc\'ia Jim\'enez and Buenaventura Andr\'es L\'opez for invaluable advice. Work partially supported by the Spanish research contracts FPA2008-00592,  FIS2008-01323, UCM-BSCH GR58/08 910309 and the FPI programme (BES-2009-013672).
 
\end{theacknowledgments}


\end{document}